\documentclass[a4paper,11pt]{article}
\usepackage{jinstpub} % for details on the use of the package, please see the JINST-author-manual
\usepackage{lineno}
\title{New developments in 3D-trench electrode sensors}

\author[a,b,1]{Jixing Ye,\note{Corresponding author.}}
\author[c,b]{Maurizio Boscardin,}
\author[c,b]{Matteo Centis Vignali,}
\author[c,b]{Francesco Ficorella,}
\author[c,b]{Omar Hammad Ali,}
\author[d]{Adriano Lai,}
\author[d]{Angelo Loi,}
\author[c]{Laura Parellada Monreal,}
\author[c,b]{Sabina Ronchin,}
\author[a,b]{Gian-Franco Dalla Betta}

\affiliation[a]{Dipartimento di Ingegneria Industriale, Università degli Studi di Trento, Via Sommarive 9, 38123 Trento, Italy}
\affiliation[b]{Trento Institute for Fundamental Physics and Applications—Istituto Nazionale di Fisica Nucleare
(TIFPA-INFN), Via Sommarive 14,38123 Trento, Italy}
\affiliation[c]{Fondazione Bruno Kessler, Via Sommarive 18,38123 Trento, Italy}
\affiliation[d]{INFN, sezione Cagliari, Dipartimento di Fisica, strada provinciale per Sestu, km 0,700, Monserrato, Italy}

\emailAdd{jixing.ye@unitn.it}

\abstract{Future high-luminosity hadron collider experiments feature unprecedented levels of event pile-up and extreme radiation environments, calling for sensors capable of 4D tracking, even after significant radiation damage. To this purpose, 3D sensors represent a viable solution, since they provide excellent radiation tolerance and very good temporal resolution. In particular, owing to the uniform electric field and weighting field distributions, 3D-trench electrode sensors from the INFN TIMESPOT project have shown a temporal resolution of $\sim$10 ps after irradiation fluences up to 1$\times$10$^{17}$ 1-Mev n$_{eq}$/cm$^2$. 
In spite of the excellent performance of these sensors, 3D-trench pixel technology is not yet fully established and the fabrication yield is not yet adequate for the production of large size pixel sensors.  
To improve the potential of the 3D-trench concept for large-area sensors, a new batch of sensors was designed at the University of Trento and fabricated at FBK, as part of the AIDA Innova project. Besides introducing some process improvements, this batch includes two different sensor variants: the standard one with continuous ohmic trenches, and a modified one with dashed ohmic trenches. On-wafer electrical test results show that most of the sensors have low leakage current and high breakdown voltage. Moreover, the fabrication yield for the new design variant is higher than that of the standard design.}


\begin{document}
\maketitle
\flushbottom

\section{Introduction}
\label{sec:intro}
    The success of the Large Hadron Collider (LHC) paved the way for the upgrade to the High Luminosity LHC (HL-LHC), which has the ability to facilitate detailed studies of known phenomena, as well as the observation of very rare new events ~\cite{HL-LHC}. To advance the measurements to the next level of precision and increase the probability of discovering new physics particles, a successor to the HL-LHC, such as the Future Circular Hadron Collider (FCC), is necessary ~\cite{FCC, Abada19}. However, the massive event pile-up and extreme radiation environments that accompany such experiments will require sensors with 4D tracking capabilities to disentangle different vertices, even after fluences up to $\sim$10$^{17}$ 1-MeV n$_{eq}$/cm$^2$.
    
    Though remarkable temporal resolutions ($\sim$30 ps) have been achieved in state-of-the-art Ultra-Fast Silicon Detectors (UFSDs) ~\cite{Arcidiacono20}, further sensor engineering is required to enhance their radiation hardness. 3D sensors, on the other hand, are known for their extreme radiation hardness due to the short inter-electrode distance. Moreover, temporal resolutions comparable to UFSDs have been reported for 3D sensors with columnar electrodes ~\cite{Kramberger19, Diehl24}, while even better performances have been found when the columnar electrodes are replaced with trenches ~\cite{Lampis23}. Most importantly, studies show that 3D sensors with trench electrodes can maintain their high temporal resolution with slightly decreased efficiency after radiation damage up to 1$\times$10$^{17}$ 1-MeV n$_{eq}$/cm$^2$ ~\cite{Addison24}.
    
    Despite the extraordinary performances in terms of temporal resolution and radiation hardness, the fabrication technology of 3D-trench electrode sensors is still in development. Defects in the long, continuous ohmic trench electrodes present in the original design \cite{Forcolin20} were observed in the TIMESPOT batches, that limited the fabrication yield. To address this issue, besides process improvements, a new design variant incorporating small gaps in the ohmic trenches was proposed. Results of TCAD and Monte-Carlo simulation show comparable temporal resolution and charge collection efficiency (CCE) of the standard and new design ~\cite{Jixing23}.
    
    Following the proof-of-concept study, a new batch of sensors including both designs was fabricated at FBK as part of the AIDA Innova project. In this paper, detailed information about this batch are presented, covering aspects from layout design to reticle arrangements. On-wafer electrical tests are also reported to provide insight into the yield of the two designs.

\section{AIDA Innova 3D-trench electrode sensors}
\subsection{Device description}   
\begin{figure}[htbp]
    \centering
    \includegraphics[width=.4\textwidth]{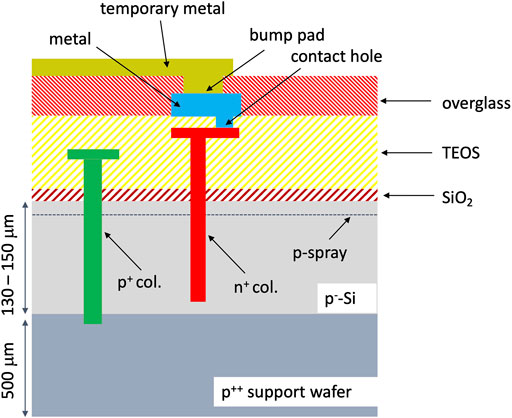}
    \qquad
    \includegraphics[width=.5\textwidth]{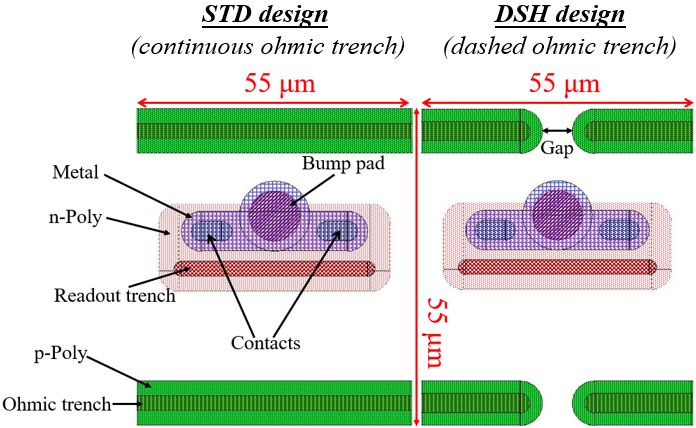}
    \caption{Schematic cross-section (left) and pixel layouts (right) of 3D-trench sensors. \label{fig:Schematic}}
    \end{figure}

    The fabrication inherited the single-sided technology that has been developed for small pitch 3D pixel sensors, in which stepper lithography is used to enhance the definition of layout details ~\cite{Boscardin21}. As indicated by Figure ~\ref{fig:Schematic} (left), the active bulk has a thickness of 150$\mu$m and is bonded to a highly doped p-type support wafer. Though all electrodes are fabricated from the front side, the ohmic trenches are made passing through the entire active volume in order to allow direct bias from the backside. The junction electrodes, on the other hand, are kept at a safe distance from the backside to prevent early breakdown. Moreover, a uniform p-spray layer is present on the front surface to isolate the pixels. Figure ~\ref{fig:Schematic} (right) shows the pixel layout of the standard (STD) and dashed (DSH) sensors. 

\subsection{Reticle arrangement}
    To explore the full potential of the reticle and understand the impact of sensor size on the yield, its layout is designed to accommodate sensors of various array sizes: 32$\times$32, 64$\times$64, and 128$\times$128. The larger sensors are intended for future Read-Out Chips (ROCs) such as the IGNITE (INFN) and Picopix (CERN), whereas the smaller ones are compatible with existing ROCs from the TIMESPOT project. The baseline pixel size is 55$\times$55 $\mu$m$^2$; smaller pixels - 42$\times$42 $\mu$m$^2$- have been introduced in test structures to start exploring the impact of smaller inter-electrode spacings on the performance.
    
\begin{figure}[htbp]
    \centering
    \includegraphics[width=.8\textwidth]{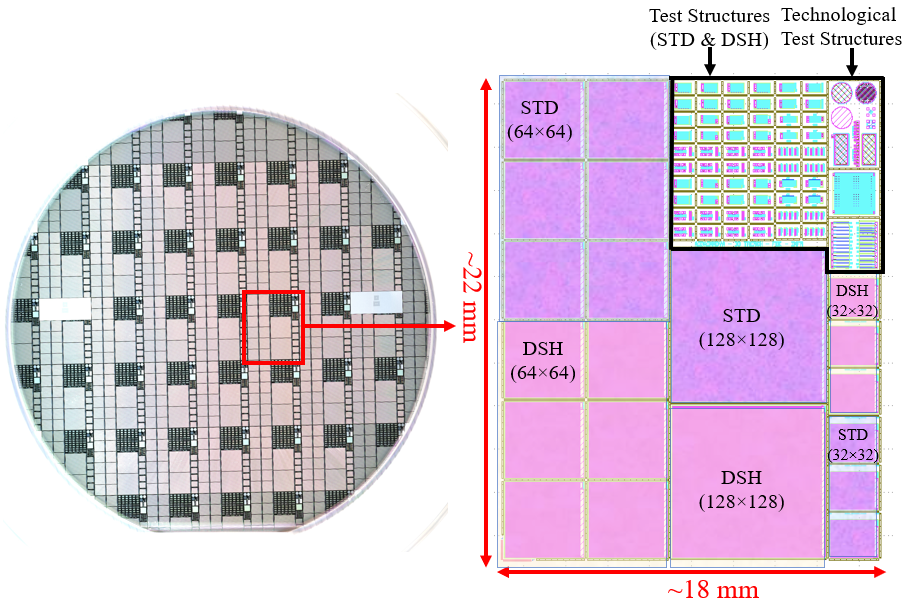}
    \caption{(Left) Photograph of a processed wafer with 29 reticle shots and (right) layout of the reticle.\label{fig:RecticleLayout}}
    \end{figure}

    As can be seen from Figure ~\ref{fig:RecticleLayout}, each reticle includes 6 STD and 6 DSH sensors with an array size of 64$\times$64, located on the left side; while on the right side sit 3 STD and 3 DSH sensors with the size of 32$\times$32. Every reticle also hosts 1 STD and 1 DSH sensors with the size of 128$\times$128. Technological test structures are located at the top right of the reticle, which enables investigation of the fabrication processes. The block of test structures located nearby includes both STD and DSH sensors with different pixel sizes, configured as groups of individual pixels, strips, and diodes. It should be stressed that, compared to the TIMESPOT batches that only had 11 reticle shots per wafer, this batch was processed with an enhanced technology to improve the yield and increase the device density, for which two options were implemented: 18 or 29 reticle shots per wafer. It is worth noting that for both wafer layout configurations the wafer bow and warp were found to be remarkably small, below 20 $\mu$m in most case and below 60 $\mu$m in the three worst wafers.

\section{Electrical characterization}
    On-wafer measurements were performed at room temperature and in dark conditions using the automatic probe station available at FBK, which can provide a maximum bias voltage of 100 V. The current-voltage (I-V) curves of all pixel sensors were tested using a temporary metal \cite{Giacomini13}. 
    
\begin{figure}[htbp]
    \centering
    \includegraphics[width=.45\textwidth]{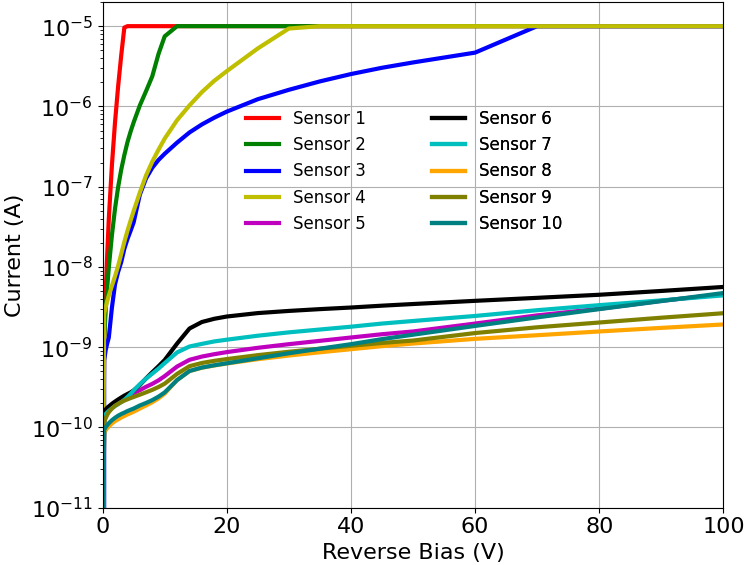}
    \qquad
    \includegraphics[width=.45\textwidth]{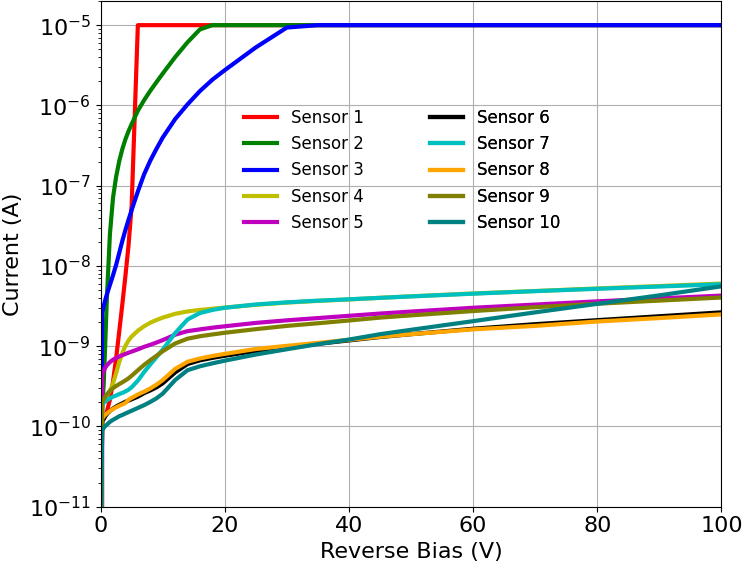}
    \caption{Leakage current of different designs with the array size of 128$\times$128: STD (left) and DSH (right).\label{fig:LeakageCurrent}}
    \end{figure}

    As an example, Figure ~\ref{fig:LeakageCurrent} shows the I-V curves of two groups of ten 128$\times$128 sensors for both designs. It can be seen that there are sensors from both designs with early breakdown, indicating the presence of defects. However, in good sensors the leakage currents are very small (well below 1 pA per pixel, after normalization) and the breakdown voltages are higher than 100 V, which is promising for achieving full depletion after strong radiation damage. 
    
\begin{figure}[htbp]
    \centering
    \includegraphics[width=.45\textwidth]{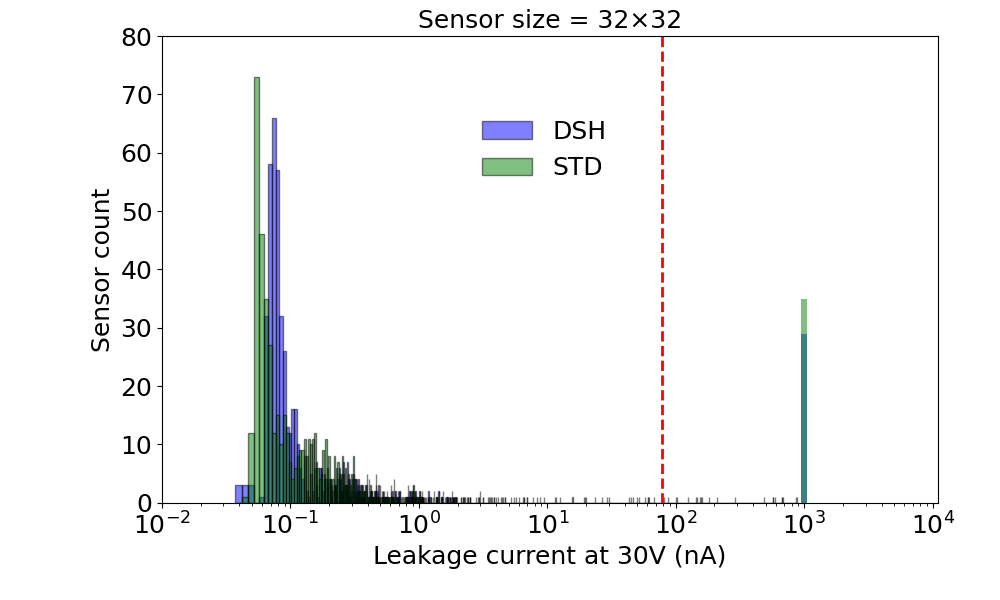}
    \qquad
    \includegraphics[width=.45\textwidth]{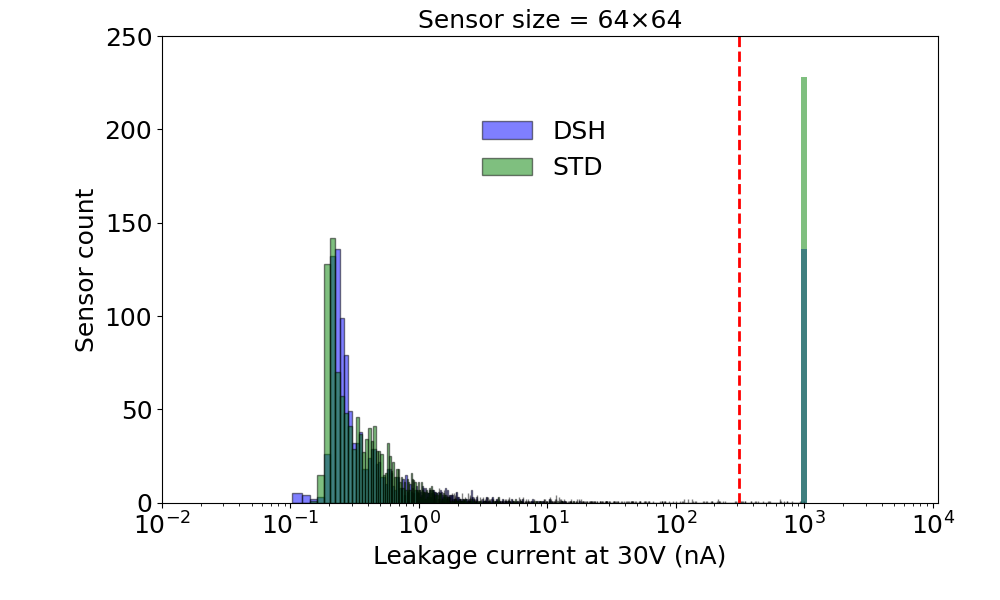}
    \qquad
    \includegraphics[width=.45\textwidth]{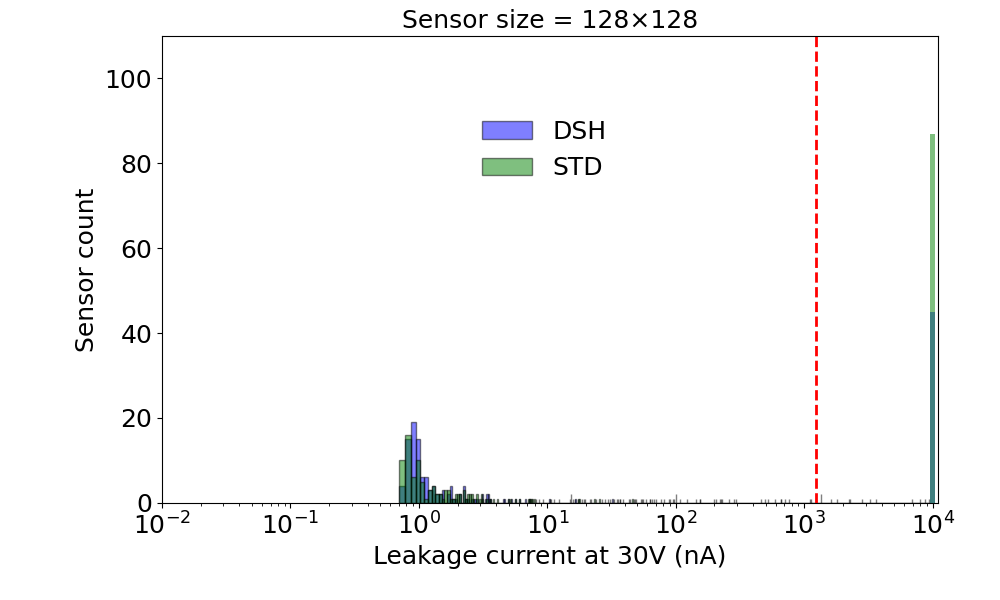}
    \caption{Distribution of the leakage current at 30 V reverse bias in different pixel arrays: 32$\times$32 (top left), 64$\times$64 (top right) and 128$\times$128 (bottom). The red dashed line represents the current limit for ATLAS ITk specifications.\label{fig:Yield}}
    \end{figure}

Similar results were obtained for sensors with the array size of 64$\times$64 and 32$\times$32. However, the fraction of sensors showing early breakdown was found to be decreasing as the total size is decreasing. This can be clearly appreciated  from Figure ~\ref{fig:Yield}, which shows the leakage current distribution at 30 V reverse bias for pixel designs with arrays sizes of 32$\times$32, 64$\times$64 and 128$\times$128. In all sub-figures, as a reference, the leakage current limit set by ATLAS ITk pixel specifications (i.e., 2.5 $\mu$A/cm$^2$) is highlighted with the dashed red lines. The decreasing trend of the yield of good sensors with the total size is evident.

For all sensor sizes, but particularly for the larger ones, it can also be observed that the DSH design shows a smaller fraction of sensors with high leakage current. This supports the initial assumption that dashed ohmic trenches can help reducing the number of defects. 

Finally, it should also be noted that the yield of good sensors was found to be independent from the wafer layout device density, proving the reliability of very dense wafer layout. 

\section{Conclusions}
    3D-trench electrode sensors from the INFN TIMESPT project demonstrated excellent temporal resolution and high radiation tolerance, but their fabrication technology is not yet established. Among possible options aimed at improving the yield from the sensor design point of view, the dashed ohmic trench concept has been proposed. Funded by the AIDAInnova project, a new batch of sensors containing both the continuous and dashed ohmic trench designs was fabricated at FBK. 
    
    By virtue of the improved fabrication process, very high wafer layout densities were successfully implemented, up to 29 reticle shots per wafer, while keeping the wafer bow and warp under control. Preliminary results from leakage current tests show very high yields (>90\%) for the 32$\times$32 sensors in both design variants. Albeit still better than for the previous TIMESPOT batch, the yield for larger sensor arrays (64$\times$64 and 128$\times$128) decreases as the total size is increased, calling for further technology optimization. However, a sizably larger fraction of good sensors was found in the dashed trench version, proving the effectiveness of the new design in reducing the defect density.
    
    Functional measurements are under way on test structures to assess the timing performance. Moreover, the best six wafers were selected to proceed with bump bonding and flip chipping to readout chips.

\appendix
\acknowledgments
    The authors acknowledge funding from INFN-CSN5 with Project TimeSPOT, and the European Union's Horizon 2020 Research and Innovation programme under GA no. 101004761.

% Bibliography

%% [A] Recommended: using JHEP.bst file
\bibliographystyle{JHEP}
%%\bibliography{biblio.bib}

%% or
%% [B] Manual formatting (see below)
%% (i) We suggest to always provide author, title and journal data or doi:
%% in short all the informations that clearly identify a document.
%% (ii) please avoid comments such as "For a review'', "For some examples",
%% "and references therein" or move them in the text. In general, please leave only references in the bibliography and move all
%% accessory text in footnotes.
%% (iii) Also, please have only one work for each \bibitem.

\end{document}